\documentclass[conference]{IEEEtran}
\IEEEoverridecommandlockouts
\usepackage{cite}
\usepackage[switch]{lineno}
\usepackage{amsmath,amssymb,amsfonts}
\usepackage{algorithm}
\usepackage{algorithmic}
\usepackage{graphicx}
\usepackage{textcomp}
\usepackage{xcolor}
\usepackage{url}
\usepackage{array} 
\usepackage{float}
\usepackage{booktabs}
\usepackage{listings}
\usepackage{changepage}
\usepackage[newfloat, frozencache, cachedir=minted-cache]{minted}
\usepackage{setspace}
\usepackage{caption}
\usepackage{pifont}
\usepackage{tcolorbox}

\makeatletter
\let\NAT@parse\undefined
\makeatother
\usepackage{hyperref}

\usepackage{multirow}

\newcolumntype{P}[1]{>{\ttfamily\arraybackslash}p{#1}}

\begin{document}
\title{Fix the Tests: Augmenting LLMs to Repair Test Cases with Static Collector and Neural Reranker}

\DeclareRobustCommand{\authorrefmark}[1]{\(^{#1}\)}
\DeclareRobustCommand\CorrespondingAuthor{\S}
\author{
\IEEEauthorblockN{
    Jun Liu\authorrefmark{1, 3, \dag},
    Jiwei Yan\authorrefmark{2, \CorrespondingAuthor, \ddag},
    Yuanyuan Xie\authorrefmark{1, 4, \dag},
    Jun Yan\authorrefmark{1, 2, 3, \dag}
    and
    Jian Zhang\authorrefmark{1, 3, 4 \CorrespondingAuthor, \dag}
}
\thanks{\authorrefmark{\CorrespondingAuthor}Corresponding authors}
\IEEEauthorblockA{
    \authorrefmark{1}Key Laboratory of System Software (Chinese Academy of Sciences) and State Key Laboratory of Computer Science,\\ Institute of Software, Chinese Academy of Sciences  \\
}
\IEEEauthorblockA{
    \authorrefmark{2}Technology Center of Software Engineering,
    Institute of Software, Chinese Academy of Sciences \\
}
\IEEEauthorblockA{
    \authorrefmark{3}University of Chinese Academy of Sciences (UCAS) \\
}
\IEEEauthorblockA{
    \authorrefmark{4}School of Intelligent Science and Technology, Hangzhou Institute for Advanced Study, UCAS \\
}
\IEEEauthorblockA{
    Email:
    \authorrefmark{\dag}\{liuj2022, xieyy, yanjun, zj\}@ios.ac.cn,
    \authorrefmark{\ddag}yanjiwei@otcaix.iscas.ac.cn
}
}

\maketitle
\thispagestyle{plain} 

\begin{abstract}

During software evolution, it is advocated that test code should co-evolve with production code. 
In real development scenarios, test updating may lag behind production code changing, which may cause compilation failure or bring other troubles. Existing techniques based on pre-trained language models can be directly adopted to repair obsolete tests caused by such unsynchronized code changes, especially syntactic-related ones. However, the lack of task-oriented contextual information affects the repair accuracy on large-scale projects. Starting from an obsolete test, the key challenging task is precisely identifying and constructing Test-Repair-Oriented Contexts (TROCtxs) from the whole repository within a limited token size.

In this paper, we propose \textsc{Synter} (\underline{SYN}tactic-breaking-changes-induced \underline{TE}st \underline{R}epair),
a novel approach based on LLMs to automatically repair obsolete test cases via precise and concise TROCtxs construction.
Inspired by developers' programming practices, we design three types of TROCtx: class context, usage context, and environment context. Given an obsolete test case to repair, \textsc{Synter} firstly collects the related code information for each type of TROCtx through static analysis techniques automatically. Then, it generates reranking queries to identify the most relevant TROCtxs, which will be taken as the repair-required key contexts and be input to the large language model for the final test repair.

To evaluate the effectiveness of \textsc{Synter}, we construct a benchmark dataset that contains a set of obsolete tests caused by syntactic breaking changes. The experimental results show that \textsc{Synter} outperforms baseline approaches both on textual- and intent-matching metrics. With the augmentation of constructed TROCtxs, hallucinations are reduced by 57.1\%.
\end{abstract}


\section{Introduction}
\label{sec:intro}

Software evolution is a fundamental and significant aspect of software development~\cite{chapin2001types}. For large-scale software such as \textit{Kafka}~\cite{githubTagsApachekafka}, the project frequently evolves, where a new version is released in one to two weeks on average and new commits are submitted nearly every day.
As production code usually changes during software evolution, it is crucial to maintain and co-evolve associated test code to ensure that they remain effective in validating the software's functionality~\cite{skoglund2004case},~\cite{ammann2016introduction}. 
Specifically, for software invoked as libraries, it is important to co-evolve the test code to follow the production code changes, which can help developers quickly notice the backward incompatible changes that may affect its clients~\cite{underBC}. 

To automatically co-evolve production code and test code, previous studies analyze and mine the software code to extract production-test co-evolution rules and patterns~\cite{hurdugaci2012aiding},~\cite{marsavina2014studying}. However, as real-world code changes come in a great many forms, they are hard to summarize into a small number of general patterns. 
In recent years, with the rapid advancement of machine learning and Large Language Models (LLMs)~\cite{jordan2015machine},~\cite{xu2022systematic},~\cite{wang2024survey}, many studies have utilized learning-based techniques to assist the production and test code co-evolution, including obsolete test case identification~\cite{wang2021understanding}, production-test co-evolution pair extraction~\cite{sun2023revisiting}, etc., which yielded favorable results.
To further reduce the developer's burden, researchers are also concerned with repairing obsolete test cases automatically. For example, Hu et al.~\cite{ceprot} identified and updated obsolete test methods by fine-tuning a pre-trained model initialized from CodeT5~\cite{wang2021codet5}, which is currently the SOTA approach for repairing obsolete test cases. 

For this task, though directly using learning-based techniques resulted in some positive outcomes, it still faces difficulties in complex repositories. For example, API signature changing is the most common and straightforward code-changing type during software evolution. However, focusing on this type of production change, the SOTA  work \textsc{Ceprot}~\cite{ceprot} fails to repair about three-quarters of test cases on real-world projects (refer to Table~\ref{tab:humanres} in Section~\ref{sec:rq1}), which means there is still some gap before existing learning-based approaches can be applied in practice. 

To fill this gap, we target repairing signature-related code changes using the power of LLMs. Here, the signature-based code changes are also called \textbf{Syntactic Breaking Changes (SynBCs)} as they may lead to compilation errors if associated tests are not co-evolved. 
According to our investigation, accurately repairing synBC-related obsolete test cases based on LLMs faces the following key challenges.
\begin{itemize}
    \item[\textbf{\textit{C1:}}] \textbf{The repair-oriented code contexts are unclear.}
    When developers manually repair a test case, not only the signature changes of the focal method but also many other related code contexts are considered for better understanding. When fixing tests with learning-based approaches, the repair-oriented code context must be explicitly extracted. As existing works repair test cases solely with the original and updated focal methods, they are unaware of related contexts. It is essential to determine which types of contexts are essential for accurate repair and how to extract them.
    \item[\textbf{\textit{C2:}}] \textbf{The token size of code contexts is limited.} 
    Compared to small-scale models, LLMs have demonstrated their extraordinary capabilities. However, even though they are designed with increasingly larger context windows, it is impractical to simply include the entire repository contents as input. Moreover, extra irrelevant information may bring negative effects as well, which means that the extracted code context is not the more the better. As the number of context tokens for LLM input is restricted, for all the code contexts that may have relations with the changed signature, it is required to sort and pick out the most relevant ones as input. 
\end{itemize}

To this end, we propose \textsc{Synter} (\underline{SYN}tactic-breaking-changes-induced \underline{TE}st \underline{R}epair, originally called \textsc{Synbciatr}), a novel approach for repairing obsolete test cases caused by SynBCs at the method level. To address challenge \textbf{\textit{C1}}, \textsc{Synter} designs three types of contexts, including Class Contexts (\textit{ClassCtxs}), Usage Contexts (\textit{UsageCtxs}), and Environment Contexts (\textit{EnvCtxs}) as \textbf{Test-Repair-Oriented Contexts (TROCtxs)}. These contexts focus on different aspects of changed codes to provide adequate contextual information for test repair.
For challenge \textbf{\textit{C2}}, \textsc{Synter} identifies and constructs qualified TROCtx from the repository in two stages. \ding{182} First, \textsc{Synter} collects all types of TROCtx by static code analysis, specifically, using the \textit{Language Server}~\cite{LSPweb}. \ding{183} Then, inspired by the idea of Retrieval-Augmented Generation (RAG)~\cite{zhang2024raft}, \textsc{Synter} generates reranking queries to identify the most relevant TROCtxs in each type according to the repair requirement extracted from the original test case, which is based on \textit{Neural Rerankers}~\cite{mei2014multimedia}. After constructing TROCtxs, \textsc{Synter} aggregates the test-repair-required information to a final prompt and generates the repaired test case.

To evaluate the effectiveness of \textsc{Synter}, we construct a benchmark dataset based on existing work which consists of 136 samples with diverse SynBCs. The evaluation is based on both the textual match and intent match metrics. In terms of the textual match, \textsc{Synter} achieves the best performance against baselines specifically on CodeBLEU (83.3), DiffBLEU (46.7), and Accuracy (32.4\%). In terms of the intent match, we conduct a human evaluation on verifying whether test cases are correctly repaired without changing their original intents, in which \textsc{Synter} correctly repairs 90.4\% test cases, achieving improvements of 248.6\% and 9.8\% when compared to \textsc{Ceprot} and \textsc{Naivellm} respectively. Moreover, \textsc{Synter} is capable of reducing 57.1\% hallucinations caused by \textsc{Naivellm}.

We make the following contributions in this paper:
\begin{itemize}
    \item We design three types of TROCtx to provide adequate contextual information for repairing obsolete test cases caused by SynBCs.
    \item We propose \textsc{Synter} to construct TROCtx by combining the static collector and neural reranker, which is utilized to enhance the repairing ability of naive LLM.
    \item Experimental results on the benchmark dataset demonstrate that \textsc{Synter} can repair obsolete test cases caused by SynBCs more effectively compared to both \textsc{Ceprot} and \textsc{Naivellm}.
\end{itemize}
The data and code are both publicly available at: 
\textit{\url{https://github.com/nonsense-j/SynTeR}}.

\begin{figure*}[t]
\centering
\setlength{\abovecaptionskip}{5pt}
\setlength{\belowcaptionskip}{-10pt}
\includegraphics[width=\textwidth]{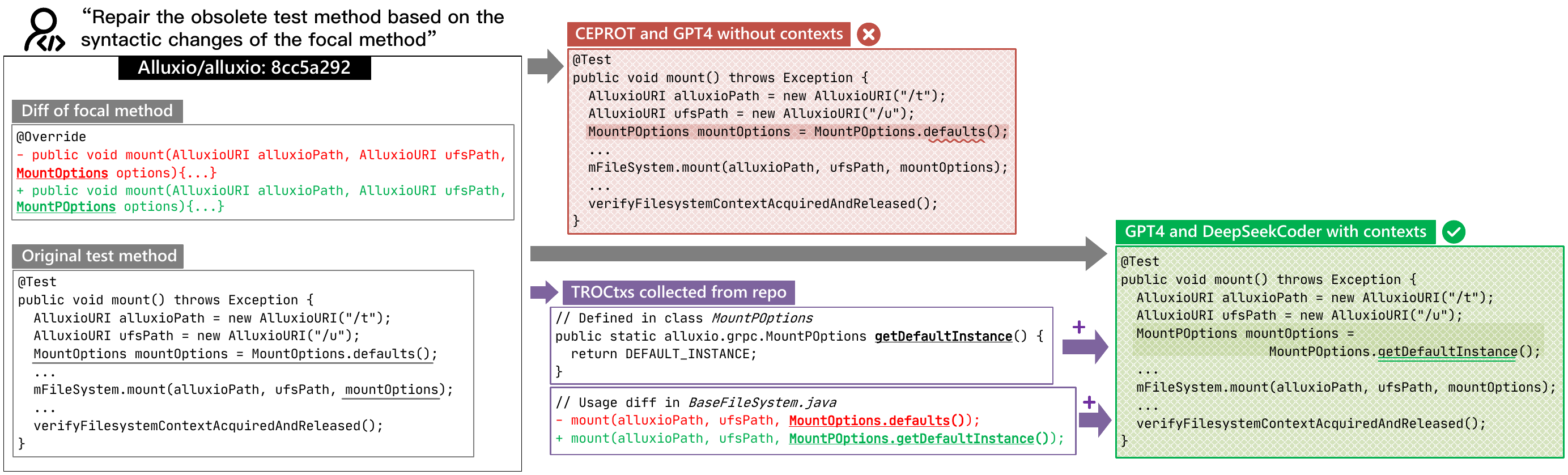}
\caption{A motivating example with a ParamSynBC collected from a commit (8cc5a292) of Alluxio/alluxio. The original test method can be correctly repaired only when we provide precise Test-Repair-Oriented Contexts (shown as TROCtxs) to LLMs.}
\label{fig:case}
\end{figure*}

\section{Background and Motivation}

\subsection{Task Definition}
\label{sec:taskdef}
Referring to previous studies~\cite{underBC},~\cite{zhang2022has},~\cite{JourJava}, Syntactic Breaking (SynB) issues represent signature-based compilation errors in API evolution, such as \textit{ClassNotFoundException} and \textit{NoSuchMethodError}. In this paper, we name method signature changes that may cause SynB issues as Syntactic Breaking Changes (SynBCs). 

We define the signature of a method (\textbf{Method Signature}) as a 5-tuple $\mathit{ms} = \langle n, P, r, M, E \rangle$, where:
\begin{itemize}
  \item $n$ is the name of the method;
  \item $P$ is a list of parameter types;
  \item $r$ is the type of the return value;
  \item $M$ is a set of modifiers;
  \item $E$ is a set of exception types that can be thrown.
\end{itemize}

Given a method that changes from $m$ to $m'$, we define a \textbf{ Syntactic Breaking Change (SynBC)} as:
$$m \xrightarrow{\mathrm{SynBC}} m'\quad \mathrm{iff}\ m.\mathit{ms} \neq m'.\mathit{ms},$$
where $m.\mathit{ms}$ and $m'.\mathit{ms}$ denote the method signatures before and after the change respectively. To determine whether SynBCs are common in production-test co-evolution, we conducted an empirical study to count the SynBCs in the breaking change dataset released by \textsc{Ceprot}~\cite{ceprot}. Each sample in the dataset contains a change of focal method on a real-world commit from GitHub. It reveals that signature-based focal changes occur in over 40\% samples of the dataset, in which parameter and return type-related changes account for a large proportion.

Specifically, we further categorize SynBCs into the following three types according to the changed elements in the method signature.
Note that, a SynBC whose parameter and return types in the method signature both change is a ParamSynBC and RetSynBC at the same time.
\begin{itemize}
    \item \textbf{Parameter-related Syntactic Change (ParamSynBC)}. 
    For a SynBC from $m$ to $m'$, if the parameter type list in the method signature changes, it is a ParamSynBC.
    $$m \xrightarrow{\mathrm{ParamSynBC}} m'\quad \mathrm{iff}\  m.\mathit{ms}.P \neq m'.\mathit{ms}.P$$
    \item \textbf{Return-related Syntactic Change (\textbf{RetSynBC})}. 
    For a SynBC from $m$ to $m'$, if the type of the return value in the method signature changes, it is a RetSynBC.
    $$m \xrightarrow{\mathrm{RetSynBC}} m'\quad \mathrm{iff}\  m.\mathit{ms}.r \neq m'.\mathit{ms}.r$$
    \item \textbf{Normal Syntactic Change (\textbf{NormSynBC})}. 
    For a SynBC from $m$ to $m'$, if it does not belong to ParamSynBC or RetSynBC type, it is a NormSynBC.
    \begin{align}
    &m \xrightarrow{\mathrm{NormSynBC}} m' \quad  \mathrm{iff}\  m.\mathit{ms} \neq m'.\mathit{ms}\ \wedge \notag\\
    &\qquad\quad m.\mathit{ms}.P = m'.\mathit{ms}.P \wedge m.\mathit{ms}.r = m'.\mathit{ms}.r \notag
    \end{align}
\end{itemize}

Here, for a SynBC that happens between the original focal method $m$ and the updated focal method $m'$, $\it{ro}'$ denotes the repository code that the method $m'$ belongs to, and $t$ denotes the obsolete test case associated with $m$. Based on this knowledge, the test-repair task hopes to get the repaired test case $t'$ as the final output. The whole task consists of two main steps. 
First, we extract the repair-oriented contexts for test $t$ from repository $\it{ro}'$ according to the SynBC from $m$ to $m'$. That process can be denoted as:
$$Construct(m, m', \it{ro}', t) = C \label{eq:construct},$$
where $C$ represents the constructed contexts. 
Then, we repair the obsolete test with the constructed contexts $C$, which can be expressed as:
$$Repair(m, m', C, t) = t' \label{eq:repair},$$
where $t'$ denotes the final repaired test case.

\subsection{Language Server and Neural Reranker}

A \textit{Language Server} consists of programming language-specific tools like static analyzers and compilers. In modern Integrated Development Environments (IDEs), language servers provide language-specific features like `\textit{autocomplete}', `\textit{goto definition}', `\textit{find usages}', and others~\cite{agrawal2024monitor}. 
Recently, the Microsoft team has created a standard JSON-RPC-based protocol, called \textbf{Language Server Protocol (LSP)}~\cite{LSPweb}, based on which multiple IDEs can communicate with the same language server to access intelligent programming features.

A \textit{Neural Reranker} is a type of machine learning model~\cite{yin2019reranking} used to reorder a given set of documents based on their relevance to a given query, which is the initial request for information expressed as keywords or complex expressions~\cite{mei2014multimedia}. It is widely used in research fields such as information retrieval, natural language processing, and recommendation systems. Instead of ranking based on simple heuristics like the frequency of query terms appearing in the query, neural rerankers are trained to take into account more complex features, like the semantic similarity between the query and the documents.

\subsection{Motivating Example}
The example in Fig.~\ref{fig:case} is used to demonstrate our motivation. It was collected from a real commit in project \textit{Alluxio}~\cite{Alluxio}. As shown in the given commit, there is a ParamSynBC for the focal method (named \texttt{mount}), where the third parameter changes from \texttt{MountOptions} to \texttt{MountPOptions}. If the associated original test case (also named \texttt{mount}) does not co-evolve with the change of the focal method, the test will fail for compilation errors (\textit{cannot resolve type}).

To automatically repair the obsolete test method, we first apply existing learning-based techniques directly. However, both \textsc{Ceprot} and GPT-4 fail due to using an undefined method (\texttt{MountPOptions.defaults()}) to construct the third parameter, where the correct method invocation should be \texttt{MountPOptions.getDefaultInstance()}. This hallucination occurs for lacking Test-Repair-Oriented Contexts (TROCtxs) in the input. After providing the required contexts shown in Fig.~\ref{fig:case}, by including either the definition of \texttt{getDefaultInstance} in class \texttt{MountPOptions} or the usage change of focal method in other production code, LLMs can generate the correct repaired test. 

When developers are asked to repair test for this case in IDEs, it is convenient for them to refer to the related contexts (TROCtxs in Fig.~\ref{fig:case}) with the programming features (such as \textit{`goto definition'} and \textit{`find usages'}) provided by language servers. Inspired by these practices, \textsc{Synter} constructs TROCtxs by simulating developers' behaviors in IDEs. Specifically, \textsc{Synter} collects related contexts by interacting with the language server and filters out unnecessary ones based on neural rerankers. Finally, \textsc{Synter} uses LLMs to repair obsolete test cases with constructed TROCtxs.

\section{Methodology}
In this section, we first show the overall framework of \textsc{Synter}. Then we demonstrate all the types of TROCtxs identified by \textsc{Synter}. Finally, we introduce the technical details of the main modules in \textsc{Synter}.

\begin{figure}[!b]
\centering
\setlength{\abovecaptionskip}{5pt}
\setlength{\belowcaptionskip}{-5pt}
\includegraphics[width=0.45\textwidth]{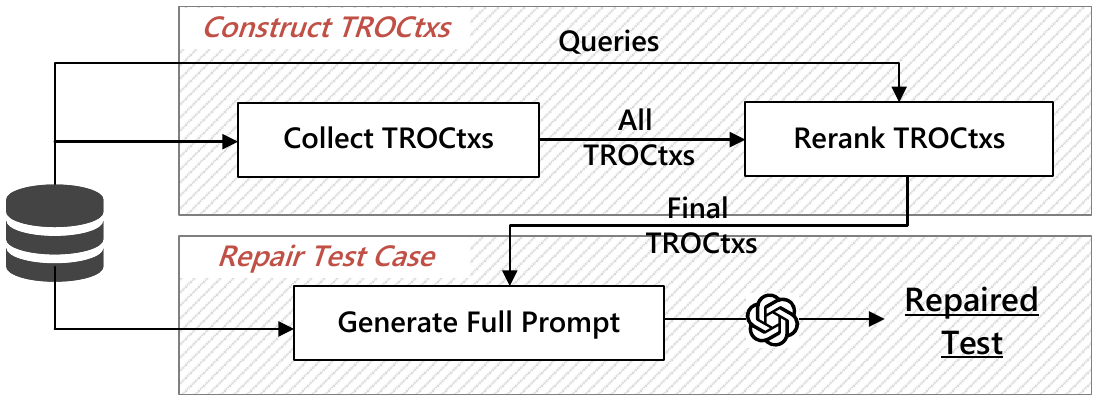}
\caption{Overall pipeline of \textsc{Synter}.}
\label{fig:struct}
\end{figure}
\subsection{Overview}
\begin{figure*}[t]
\centering
\setlength{\abovecaptionskip}{5pt}
\setlength{\belowcaptionskip}{-10pt}
\includegraphics[width=\textwidth]{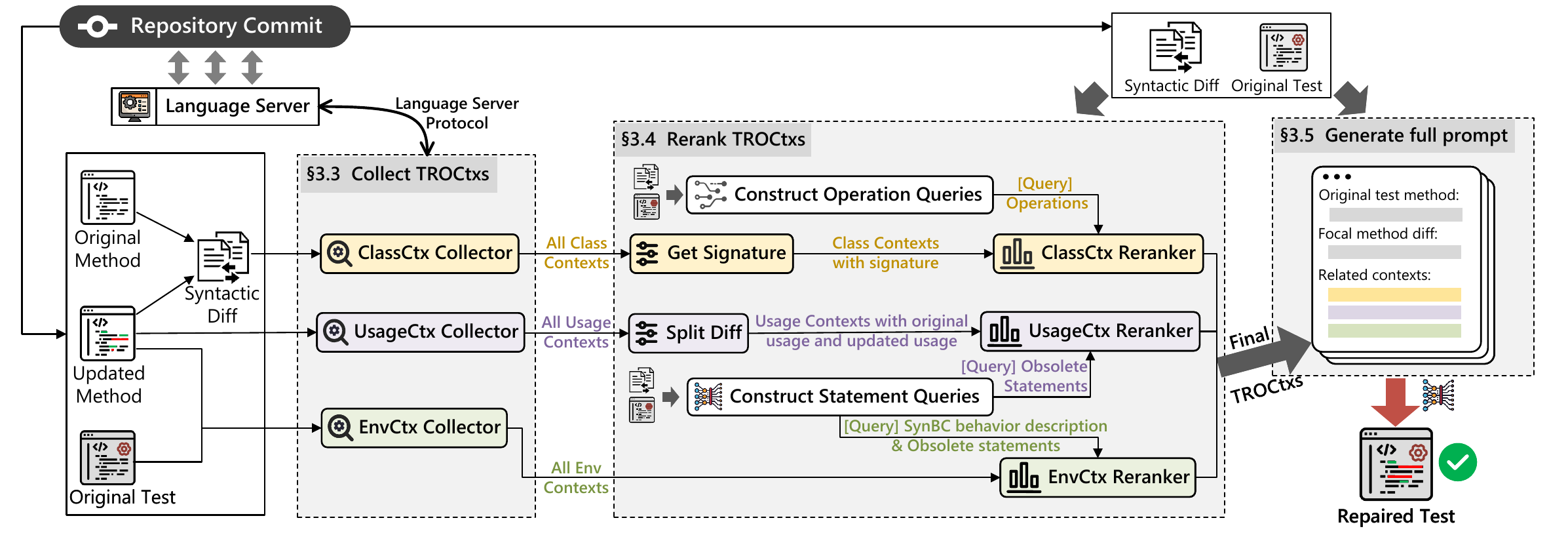}
\caption{Overview of \textsc{Synter}.}
\label{fig:overview}
\end{figure*}

The overall pipeline of \textsc{Synter} is depicted in Fig.~\ref{fig:struct}. 
Given the change of the focal method and the obsolete test method as inputs, \textsc{Synter} consists of three major steps. \textbf{(1) Collecting TROCtxs}: \textsc{Synter} analyzes all the related contexts from inputs and requests the language server to collect and process them into candidate chunks; \textbf{(2) Reranking TROCtxs}: \textsc{Synter} reranks candidate chunks with queries constructed from the inputs; \textbf{(3) Generating full prompt}: \textsc{Synter} aggregates the inputs and final TROCtxs to generate the full prompt, which is used to repair the test with LLM. We also provide a detailed overview in Fig.~\ref{fig:overview}.

\subsection{Types of TROCtx}
Based on the characteristics of the test-repair task, we categorize TROCtxs into three types: Class Contexts (ClassCtxs), Usage Contexts (UsageCtxs), and Environment Contexts (EnvCtxs). In alignment with real-world developers' practices, these categories of contexts can provide comprehensive and sufficient contextual information for test repair.

\begin{itemize}
    \item[$\bigstar$] \textbf{ClassCtxs} include the member accesses (method and field accesses) of a specific class and its parent classes. These contexts indicate the accurate operations supported by a given class type, serving to alleviate the hallucination of LLMs. Since new class types can be introduced in parameter types and the return type, ClassCtxs will be collected specially for ParamSynBCs and RetSynBCs. For example, the ClassCtxs for the case caused by a ParamSynBC in Fig.~\ref{fig:case} are partly demonstrated in Listing~\ref{lis:classctx}.

\begin{listing}[!ht]
\begin{minted}
[
frame=single,
framesep=2mm,
baselinestretch=1,
fontsize=\scriptsize,
breaklines=true
]
{java}
// defined in MountPOptions and its parent classes
// methods are simplified as signatures
...
public static final int READONLY_FIELD_NUMBER = 1;
public boolean hasReadOnly();
public static MountPOptions getDefaultInstance();
...
\end{minted}
\caption{ClassCtxs of the new class \texttt{MountPOptions}.}
\label{lis:classctx}
\end{listing}
    
    \item[$\bigstar$] \textbf{UsageCtxs} include the changes of usages for the focal method in the diff format. Usages of the updated focal method in other parts of the repository can illustrate how to properly call the method in the associated test. UsageCtxs will be collected for all the types of SynBCs. For example, the UsageCtxs for case introduced in Fig.~\ref{fig:case} are partly demonstrated in Listing~\ref{lis:usagectx}.

\begin{listing}[!ht]
\begin{minted}
[
frame=single,
framesep=2mm,
baselinestretch=1,
fontsize=\scriptsize,
breaklines=true
]
{diff}
...
- mount(alluxioPath, ufsPath, MountOptions.defaults()
+ mount(alluxioPath, ufsPath, MountPOptions.getDefaultInstance()
...
\end{minted}
\caption{UsageCtxs of the focal method \texttt{mount()}.}
\label{lis:usagectx}
\end{listing}

    \item[$\bigstar$] \textbf{EnvCtxs} include the environmental changes of the focal and test method in the diff format. Given a method $m$, we define the class containing it and its parent classes as the environment of $m$. Only code changes external to $m$ in the environment will be collected to construct its EnvCtxs. These contexts can indicate updates of related identifiers and similar change patterns. EnvCtxs will be collected for all the types of SynBCs. For example, the EnvCtxs of the focal method for the case introduced in Fig.~\ref{fig:case} are partly demonstrated in Listing~\ref{lis:envctx}.
    
\begin{listing}[!ht]
\begin{minted}
[
frame=single,
framesep=2mm,
baselinestretch=1,
fontsize=\scriptsize,
breaklines=true
]
{diff}
// code diffs of class containing mount and its parents
...
- return openFile(path, OpenFileOptions.defaults());
+ return openFile(path, OpenFileOptions.getDefaultInstance());
...
\end{minted}
\caption{EnvCtxs of of the focal method \texttt{mount()}.}
\label{lis:envctx}
\end{listing}
\end{itemize}

\textsc{Synter} aims to collect and rerank different types of TROCtxs respectively. In the following subsections, we will describe the details of how each type of TROCtx is collected, reranked, and ultimately aggregated as displayed in Fig.~\ref{fig:overview}.

\subsection{Collecting TROCtxs}
To collect repo-level contexts, \textsc{Synter} interacts with language servers via \textit{LSP}~\cite{LSPweb}. Every request message conforms to the JSON-RPC-based protocol, which consists of the request type (such as `\textit{goto definition}') and the cursor position of the request identifier, including the file path and the indices of the line and column for the identifier within the file.

For every type of TROCtx, \textsc{Synter} first automatically analyzes the inputs and locate \textbf{key identifiers} (identifiers that need to find definitions and references by LSP) with their positions using the static parser \textit{tree-sitter}\footnote{\url{https://tree-sitter.github.io/tree-sitter/}}. 
The required contexts are further collected through the language server using the python library \textit{multilspy}\footnote{\url{https://github.com/microsoft/monitors4codegen}}.

\subsubsection{\textbf{Collecting ClassCtxs}}
Based on the syntactic diff of the focal method, \textsc{Synter} first identifies the new class types introduced in the method signature of the updated version (both parameter types and the return type will be checked). These new class types are considered as key identifiers for collecting ClassCtxs. Through requesting the language server, \textsc{Synter} collects the definitions of these new class types as well as their parent classes. 
To split the collected contexts into identical chunks, \textsc{Synter} cleans up all comments and the definition body is divided into member access operations. Only non-private declarations of fields and methods are collected. Specifically, for classes and private field accesses with Lombok annotations\footnote{\url{https://projectlombok.org/features/}}(\texttt{@Data}, \texttt{@GETTER}, \texttt{@SETTER}), \textsc{Synter} also retains their related private field accesses.
 
For the case in Fig.~\ref{fig:case} as an example, the updated focal method uses a new parameter type (\texttt{MountPOptions}). \textsc{Synter} will collect all the declarations of field and method to construct ClassCtxs, as demonstrated in Listing~\ref{lis:classctx}.

\subsubsection{\textbf{Collecting UsageCtxs}}
For UsageCtxs, the name identifier of the updated focal method is considered as the key identifier. The collecting procedures are described in Algorithm~\ref{alg:collectusage}.
On line~\ref{alg:cul3}, \textsc{Synter} locates the name identifier of the updated focal in the repository.
On line~\ref{alg:cul4}, all the usages of the updated focal method are fetched by requesting the language server.
For each usage, \textsc{Synter} collects the usage-diff texts (diffs that contain the change of usage) in two steps, generating the diff (lines~\ref{alg:cul6}-~\ref{alg:cul10}) and gathering the required contexts (lines~\ref{alg:cul11}-~\ref{alg:cul18}). \textsc{Synter} formats the files that contain usages before generating diff to avoid missing information. Also, since usage-diff of invocation is not sufficient for ParamSynBCs and RetSynBCs, \textsc{Synter} collects additional backward and forward surrounding contexts respectively (lines~\ref{alg:cul13} and~\ref{alg:cul16}).


For the case in Fig.~\ref{fig:case}, \textsc{Synter} finds ten usages in the repository. After filtering out usages in comments and repeated ones, four usage-diff texts are collected, one of which is demonstrated in Listing~\ref{lis:usagectx}.

\begin{algorithm}
	\setstretch{1.1}
	\renewcommand{\algorithmicrequire}{\textbf{Input:}}
	\renewcommand{\algorithmicensure}{\textbf{Output:}}
	\small
	\caption{Algorithm of Collecting UsageCtxs}
	\label{alg:collectusage}
	\begin{algorithmic}[1]
		\REQUIRE $\sf m_{ori}$: original focal method, $\sf m_{upd}$: updated focal method,
		$\sf p$: focal relative path, $\sf lsp$: object interacting with the language server, $\sf repo$: object interacting with the repository commit
		\ENSURE $\sf UsageCtxs$: a set of usage contexts in diff format
		\STATE \emph{Initialize $\sf UsageCtxs$ as an empty set}
		\STATE $\sf synbc \gets \textit{getSynBC}(m_{ori}, m_{upd})$ \\
		\textcolor{blue}{\COMMENT{get the syntactic changes of the focal method}}
		\STATE $\sf pos \gets \textit{getMethodNamePos}(repo.\textit{getSrcFile}(p), m_{upd})$ \label{alg:cul3}\\
		\textcolor{blue}{\COMMENT{get the cursor position of the name identifier in $\sf m_{upd}$}}
		\STATE $\sf usages \gets lsp.\textit{requestUsages}(p, pos)$ \label{alg:cul4}\\
		\textcolor{blue}{\COMMENT{request the language server for usages of $\sf m_{upd}$}}
		\FORALL{$\sf usage$ \textbf{in} $\sf usages$}
		\STATE $\sf uf_{ori} \gets repo.\textit{getSrcFile}(usage.relpath)$ \label{alg:cul6}
		\STATE $\sf uf_{upd} \gets repo.\textit{getTgtFile}(usage.relpath)$ \\
		\textcolor{blue}{\COMMENT{get the original and updated files that contain usage}}
		\STATE $\sf uff_{ori} \gets \textit{format}(uf_{ori})$
		\STATE $\sf uff_{upd}, pos_{fmt} \gets \textit{formatWithCursor}(uf_{upd}, pos)$ \\
		\textcolor{blue}{\COMMENT{format the original and updated files ($\sf pos$ also updates)}}
		\STATE $\sf udiffs \gets \textit{generateDiff}(uff_{ori}, uff_{upd})$ \label{alg:cul10} \\
		\STATE $\sf utext \gets \textit{collectInvokeStmt}(udiffs, pos_{fmt})$ \label{alg:cul11}\\
		\textcolor{blue}{\COMMENT{initialize usage-diff text with invocation statement}}
		\IF{$\sf synbc $ contains change in parameter types}
		\STATE $\sf  utext \gets \textit{collectBeforeCtx}(udiffs, pos_{fmt})+utext$ \label{alg:cul13}
		\ENDIF
		\textcolor{blue}{\COMMENT{enrich usage-diff text with contexts before invocation}}
		\IF{$\sf synbc $ contains change in return type}
		\STATE $\sf  utext \gets utext + \textit{collectAfterCtx}(udiff, pos_{fmt})$ \label{alg:cul16}
		\ENDIF
		\textcolor{blue}{\COMMENT{enrich usage-diff text with contexts after invocation}}
		\STATE $\sf UsageCtxs.\textit{add}(utext)$ \label{alg:cul18}
		\ENDFOR
	\end{algorithmic}  
\end{algorithm}

\subsubsection{\textbf{Collecting EnvCtxs}}
EnvCtxs represent additional contexts indicating environmental changes, namely env-diff texts. \textsc{Synter} collects EnvCtxs for the focal and test method respectively. Specifically, the class file to which the method belongs and its parent classes are treated as the method's environment.
Here, the class name identifiers are the key identifiers, which are used to find out the related parent classes and collect their diffs as env-diff texts. Also, diffs are generated after cleaning comments and reformatting.

For the case in Fig.~\ref{fig:case}, the environment of the test method remains unchanged. Therefore, only environmental changes of the focal method are collected as env-diff texts to constructed EnvCtxs, which are partly shown in Listing~\ref{lis:envctx}.

\subsection{Reranking TROCtxs}
\textsc{Synter} utilizes neural rerankers to filter and retain the most relevant contexts. As the test cases should be functionally consistent before and after the evolution, they have high similarity. Thus, our primary idea of rerankers is to make use of the original test to construct queries for TROCtx reranking. For the case in Fig.~\ref{fig:case}, from the collected ClassCtxs of the new class \texttt{MountPOptions}, we hope to filter out unrelated member accesses and precisely retain the required method declaration ``\texttt{MountPOptions.getDefaultInstance()}''. Given the text `\texttt{MountOptions.defaults()}'' as a query, we can rerank the ClassCtxs to get the most similar APIs with neural rerankers. 
As shown in Listing~\ref{lis:rerankres}, the reranking result reveals that the required API ``\texttt{MountPOptions.getDefaultInstance()}'' is ranked with the highest score under the given query.

\begin{listing}[!ht]
\begin{minted}
[
frame=single,
framesep=2mm,
baselinestretch=1,
fontsize=\scriptsize,
breaklines=true
]
{java}
// methods are simplified as signatures
public static MountPOptions getDefaultInstance(); // top1
public MountPOptions getDefaultInstanceForType(); // top2
...
\end{minted}
\caption{Reranked ClassCtxs of class \texttt{MountPOptions}.}
\label{lis:rerankres}
\end{listing}

\subsubsection{\textbf{Constructing Queries}}

Based on the above ideas, the quality of the query texts decides the reranking results.
However, as we have three types of TROCtxs but their formats (or granularity) are different, \textsc{Synter} reranks each type of TROCtxs with distinct queries instead of using a general one.
In terms of granularity for TROCtx, ClassCtx is a fine-grained one where each context is a specific member access operation within the corresponding class (Listing~\ref{lis:classctx}), while UsageCtx and EnvCtx are coarse-grained ones whose contexts directly refer to the diff texts of statements (Listings~\ref{lis:usagectx} and~\ref{lis:envctx}). Overall, \textsc{Synter} constructs two kinds of corresponding queries, operation queries and statement queries.

\textbf{Constructing operation queries}.
We define operation as member access to a given class. Operation queries are used to rerank method and field declarations in ClassCtxs for new class types. Based on our insight, the operations used in the original test can be reused as queries, namely operation queries, to rerank ClassCtxs. For new class types in ParamSynBCs and RetSynBCs, \textsc{Synter} constructs fine-grained queries by extracting operations that could potentially be accessed in the repaired test as shown in Algorithm~\ref{alg:constopq}. Specifically, \textsc{Synter} constructs operation queries for new parameter and return class types respectively.

\begin{algorithm}
	\setstretch{1.2}
	\renewcommand{\algorithmicrequire}{\textbf{Input:}}
	\renewcommand{\algorithmicensure}{\textbf{Output:}}
	\small
	\caption{Algorithm of Constructing Operation Queries}
	\label{alg:constopq}
	\begin{algorithmic}[1]
		\REQUIRE $\sf m_{ori}$: original focal method, $\sf m_{upd}$: updated focal method, $\sf t$: original test method
		\ENSURE $\sf OpQueries$: a tuple of queries consisting of operations for new parameter and return class types
		\STATE \emph{Initialize $\sf ParamOpQ, RetOpQ$ as empty sets}
		\STATE $\sf synbc \gets \textit{getSynBC}(m_{ori}, m_{upd})$
		\IF{$\sf synbc $ is a ParamSynBC}
		\FORALL{$\sf arg$ \textbf{in} $\sf \textit{getObsArgs}(m_{ori}, m_{upd})$} \label{alg:opq4}
		\STATE $\sf op \gets \textit{getSetOp}(arg)$\\
		\textcolor{blue}{\COMMENT{construct operation queries for obsolete parameters}}
		\STATE $\sf ParamOpQ.\textit{add}(op)$ \label{alg:opq6}
		\ENDFOR
		\STATE $\sf ops \gets \textit{backwardParamsOps}(t, synbc)$ \label{alg:opq8}\\
		\textcolor{blue}{\COMMENT{backward analysis to extract operations based on synbc}}
		\STATE $\sf ParamOpQ.\textit{update}(ops)$
		\ENDIF
		\IF{$\sf synbc $ is a RetSynBC}
		\STATE $\sf ops \gets \textit{forwardReturnOps}(t, synbc)$ \label{alg:opq13}\\
		\textcolor{blue}{\COMMENT{forward analysis to extract operations based on synbc}}
		\STATE $\sf RetOpQ.\textit{update}(ops_o)$
		\ENDIF
		\STATE $\sf OpQueries \gets (ParamOpQ, RetOpQ)$
		\textcolor{blue}{\COMMENT{aggregate queries}}
	\end{algorithmic}  
\end{algorithm}

To construct operation queries for collected ClassCtxs in ParamSynBCs, starting from the invocation of the focal method in the original test, \textsc{Synter} extracts backward operations related to the obsolete parameters, which are modified during the given SynBC. 
Since the obsolete parameters may be refactored to be set by new parameters, \textsc{Synter} first adds additional operation queries in the form as \texttt{set\_xxx} on lines~\ref{alg:opq4}-~\ref{alg:opq6}. Then, on line~\ref{alg:opq8}, \textsc{Synter} traverses the def-use chains~\cite{def1} of the obsolete parameters to extract directly used operations, including method accesses and field accesses. These operations are also collected as operation queries. 


To construct operation queries for collected ClassCtxs in RetSynBCs, starting from the invocation of the focal method in the original test, \textsc{Synter} applies forward propagatability analysis by traversing the references of the return object for the focal method. During the analysis, related operations of the return object are extracted as the operation queries.

For the focal method in Fig.~\ref{fig:case}, only the parameter types changed. Therefore,  \textsc{Synter} only extracts operation queries for the newly added parameter class type \texttt{MountPOptions}, which is shown as follows. 
\begin{itemize}
    \item[$\rhd$] $\sf ParamOpQ$: \{`\textit{MountOptions.defaults()}', `\textit{setOptions()}'\}
\end{itemize}

\textbf{Constructing statement queries}.
\textsc{Synter} constructs coarse-grained queries for reranking diff texts of statements. In our settings, diff texts are collected after reformatting so that every line in the diff contains a separate statement. Since we only keep changed parts in diff texts, we should extract obsolete statements from the original test as the query, namely statement query, to rerank diff texts in UsageCtxs and EnvCtxs.

As recent studies have shown that LLM specializes in code summarizing and understanding tasks~\cite{codefslr},~\cite{wei2022chain}, \textsc{Synter} uses an LLM to extract obsolete statements from the original test. 
Following existing works~\cite{nashid2023retrieval, wang2024software}, we adopt LLMs with \textit{Few-Shots Learning} and \textit{Chain-Of-Thought} prompting to identify obsolete statements with the syntactic change of the focal method and original test method as input. The LLM is first asked to summarize the syntactic differences of the focal method, and then find the obsolete test statements. Both the \textbf{SynBC behavior description} (in natural language) and \textbf{obsolete statements} (stmts) are collected as coarse-grained statement queries. Specifically, the SynBC behavior description is used as coarse queries for reranking the EnvCtxs of the focal method, while the EnvCtxs of the test method are reranked with the obsolete statements. 


For the example in Fig.~\ref{fig:case}, the extracted statement queries are demonstrated as follows. 
\begin{itemize}
    \item[$\rhd$] SynBC behavior description: ``\textit{The method mount() has been updated to accept an object of type `MountPOptions' instead of `MountOptions' as its third parameter.}''
    \item[$\rhd$] obsolete stmts: ``\textit{MountOptions mountOptions = MountOptions.defaults(); mFileSystem.mount(alluxioPath, ufsPath, mountOptions;''}
\end{itemize}

\subsubsection{\textbf{Reranking TROCtxs with Queries}}
\textsc{Synter} reranks each type of TROCtxs with different queries as shown in Fig.~\ref{fig:overview}. During a single reranking process, the top three most relevant contexts are retained by default.

To rerank ClassCtxs, first, all the constructor declarations in ClassCtxs are retained since they are necessary to construct instances for the class. Then, for other method and field declarations, \textsc{Synter} transforms method declarations in ClassCtxs into their signatures before reranking. According to the SynBC, ClassCtxs of new parameter and return class types are reranked with $\sf ParamOpQ$ and $\sf RetOpQ$ respectively. 

To rerank UsageCtxs, every usage-diff text in the UsageCtxs is divided into original usage (by removing added lines in diffs) and updated usage (by removing deleted lines in diffs). \textsc{Synter} reranks both the original and updated usages with obsolete statements as query, in which the maximum reranking score determines the relevance of the given usage-diff text.

To rerank EnvCtxs, EnvCtxs of the focal method are directly reranked with the SynBC behavior description, while EnvCtxs of the test method are reranked with the obsolete statements.

\subsection{Generating Full Prompt}
The full prompt consists of the unified diff of the focal method, the original test, and the related contexts (TROCtxs). 

Following existing works using LLM for code tasks~\cite{li2023nuances, wang2024software}, \textsc{Synter} cleans comments in codes to avoid influencing inferring program intentions. Then, \textsc{Synter} generates the unified diff of the focal method after formatting. 

The general structure of the final prompt is demonstrated in Fig.~\ref{fig:overview}. Specifically, we start by setting up the system role of LLM as an expert in Java software evolution and briefly describe the test-repair task caused by syntactic changes of the focal method. Then, we introduce the original test that needs to be repaired followed by the unified diff of the focal method. At last, we categorize TROCtxs according to their types and ask LLM to treat these contexts as references. For each group of contexts, we also provide the basic description. For the case in Fig~\ref{fig:case} as an example, we automatically generate a description (`Defined in class MountPOptions') for ClassCtxs of \texttt{MountPOptions}.

Finally, \textsc{Synter} requests the LLM with the generated full prompt to generate the repaired test.

\section{Experimental Setup}
Before the evaluation, in this section, we first introduce the construction of our benchmark dataset and then describe the baselines and metrics used in the evaluation. Finally, we will briefly show the implementation settings of our approach.

\subsection{Benchmark Datasets}
In this study, we focus on repairing obsolete test cases caused by syntactic changes in the focal method. 
To the best of our knowledge, \textsc{Ceprot} is the first and SOTA learning-based approach to co-evolve test cases at the method level~\cite{ceprot}. Therefore, we reuse and refine the dataset provided by \textsc{Ceprot}, in which all the samples are collected from the top 1,500 Java projects on GitHub by Liu et al.~\cite{liu2020automating}.

The evaluation dataset, i.e., testing dataset, of \textsc{Ceprot} contains 520 samples extracted from 128 real-world Java projects. 
Based on that, we first filter out the samples without syntactic changes in the focal method. After this step, 211 samples remain.
Then, to ensure the high quality of the dataset, we manually \textit{filter noisy samples} and \textit{augment new samples} referring to the corresponding repository commit. 

\textbf{\textit{Filtering noisy samples}}. Samples with the following characteristics are filtered out as noises in the dataset:
\begin{itemize}
    \item the test method is incomplete or not yet implemented;
    \item the focal method is incomplete with only the signature;
    \item the focal method is not used in the given test.
\end{itemize}
We manually resolve the problems above by extracting correct co-evolution pairs with SynBCs from the repository.

\textbf{\textit{Augmenting new samples}}. To improve the diversity and generality of the dataset, we mine new samples to enrich the dataset. Specifically, we augment at most two samples from the real-world repository with different changing patterns for each commit in the original dataset.

\begin{figure}[!ht]
\centering
\setlength{\abovecaptionskip}{5pt}
\setlength{\belowcaptionskip}{-5pt}
\includegraphics[width=0.45\textwidth]{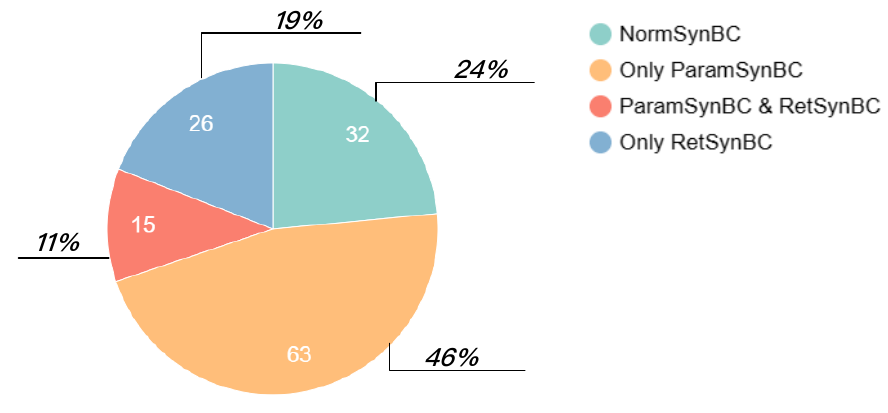}
\caption{Distribution of samples based on the type of SynBCs.}
\label{fig:dataset}
\end{figure}

To sum up, the final benchmark dataset consists of 137 samples from 54 projects. As shown in Fig.~\ref{fig:dataset}, the samples are diverse in the types of associated SynBCs.

\subsection{Baselines}
To assess the effectiveness of \textsc{Synter}, we consider two baselines from existing studies and our investigation, namely \textsc{Ceprot} and \textsc{Naivellm}.

\textbf{\textsc{Ceprot}} is the SOTA learning-based approach in updating obsolete test cases. According to Hu et al.~\cite{ceprot}, \textsc{Ceprot} is built on a code language model fine-tuned from CodeT5 and outperforms previous techniques. Taking notes that \textsc{Ceprot} needs edit sequences of the focal method as input, we reproduce it with \textit{clang-format}\footnote{\url{https://clang.llvm.org/docs/ClangFormat.html}} and \textit{difflib}\footnote{\url{https://docs.python.org/zh-cn/3/library/difflib.html}}. 
With the replication package of \textsc{Ceprot}, we retrained the model and saved the best checkpoint with the highest F1. According to the result of our replication (accuracy: 11.9\%), the performance is consistent with the statistics in the paper (accuracy: 12.3\%)~\cite{ceprot}.

\textbf{\textsc{Naivellm}} is developed based on LLM without related contexts. Compared with CodeT5, LLM is trained on huge sets of data with a much larger number of parameters. Therefore, LLM is more intelligent in general code tasks. In \textsc{Naivellm}, we directly use LLM to repair the test case with the focal diff (the unified diff of original and updated focal methods) and the original test method as input. Following the pre-processes in \textsc{Synter}, comments in codes are cleaned and all the filtered codes are reformatted to a standardized style.

\subsection{Metrics}
Based on previous studies~\cite{ceprot}~\cite{zhou2023codebertscore}, we design metrics to measure the quality of repaired test cases from two aspects, and finally get five metrics. 

\textbf{\textit{Textual Match}.} We use three specific metrics to measure how the generated code is close to the ground truth. \textbf{(1) CodeBLEU.} Developed from the classical machine translation evaluation metric \textit{BLEU}~\cite{papineni2002bleu},  \textit{CodeBLEU}~\cite{ren2020codebleu} is widely used in evaluating code generation tasks by measuring the similarity with code semantics. Thus, we use CodeBLEU to assess the similarity between the generated method and the ground truth.
\textbf{(2) DiffBLEU.} 
As only a part of statements are modified, we design the metric \textit{DiffBLEU} to concentrate on the modified ones specifically. To compute it, we calculate the BLEU score of the repaired test whose unchanged statements are removed. DiffBLEU serves as a complement to CodeBLEU. \textbf{(3) Accuracy (\textit{\%}).} This metric represents the percentage of samples where the generated method is identical to the ground truth (exactly match textually). 

\textbf{\textit{Intent Match}.} 
Besides textual match metrics, we also design two other specific metrics to measure whether the generated repaired test cases are correct.
Ideally, test repair should not change the intent of the test case, which means that the repaired test should keep the same assertions and input values. Referring to the original test and the ground truth respectively, we use two specific metrics to evaluate the \textit{repairability}. \textbf{(1) $\text{Repairability}_\text{\textit{ori}}$ (\%).} $\text{Repairability}_\text{\textit{ori}}$ represents the repairability referring to the original test. Specifically, we calculate $\text{Repairability}_\text{\textit{ori}}$ as the percentage of samples where the generated test can be successfully compiled and shares the same assertions and input values with \textit{the original test}. \textbf{(2) $\text{Repairability}_\text{\textit{gt}}$ (\%).} Similar to $\text{Repairability}_\text{\textit{ori}}$, $\text{Repairability}_\text{\textit{gt}}$ is calculated referring to the ground truth. For each repaired test case, two developers with more than three years of Java programming experience are asked to perform manual evaluation separately. If they can't reach an agreement, they will have a discussion on uncertain samples until they agree on consistent conclusions.

In addition, we also record the \textbf{Syntax Pass Rate (SPR)} and \textbf{Compilation Pass Rate (CPR)} as the success rate of syntax checking and compilation for generated codes respectively.

\subsection{Implementation Settings}
\textsc{Synter} is built based on LangChain~\cite{langchain}, which is a framework designed to simplify the procedures of developing applications powered by LLMs. With the APIs provided by LangChain, \textsc{Synter} utilizes the SOTA LLM (\textit{GPT-4})~\cite{gpt4} developed by OpenAI and the widely-used open-source neural reranker (\textit{bge-reranker-v2-m3})~\cite{bge_m3} released by BAAI~\cite{BAAI}. To reduce the impact of randomness, for each case in the dataset, we request LLM with the temperature as 0.1 three times and keep the best one for both \textsc{Naivellm} and \textsc{Synter} in evaluation.

\section{Evaluation}
Based on the constructed benchmark dataset, we evaluate the effectiveness of our approach and conduct a comprehensive analysis of the results. We address these research questions:
\begin{itemize}
    \item \textbf{RQ1:(\textit{Effectiveness of \textsc{Synter}})} Can \textsc{Synter} effectively repair obsolete tests caused by SynBCs?
    \item \textbf{RQ2:(\textit{Effectiveness of TROCtx)}} To what extent do the TROCtxs contribute to correctly repairing the test?
    \item \textbf{RQ3:(\textit{Failure Analysis)}} Under which cases does \textsc{Synter} fail to repair?
    \item \textbf{RQ4:(\textit{Efficiency})} What is the efficiency of \textsc{Synter}?
\end{itemize}

\subsection{RQ1: Effectiveness of \textsc{Synter}}
\label{sec:rq1}
\subsubsection{Basic Evaluation on Effectiveness}
To evaluate the performance of \textsc{Synter} compared with baselines, we adopt different approaches to repair obsolete test cases in the benchmark dataset. 

First, we conduct \textit{syntax validation} on the generated test codes. As shown in Tab.~\ref{tab:basicres}, the Syntax Pass Rate (SPR) of \textsc{Ceprot} is 47.8\%, while LLM-based approaches all pass the validation (100\%), which means that LLM demonstrates a higher proficiency in producing syntactically accurate code than CodeT5. 
Then we use the three textual metrics to measure the average similarity between the generated test code and the ground truth. The column \textbf{\textit{Textual Match}} in Tab.~\ref{tab:basicres} demonstrates that our approach outperforms baselines in terms of all three metrics. Specifically, LLM-based approaches have average improvements of 12.4\% and 72.4\% in terms of CodeBLEU and DiffBLEU compared with \textsc{Ceprot}, which indicates that LLM is more capable of understanding the semantics of code and generating repaired tests. With the highest scores on CodeBLEU and DiffBLEU, \textsc{Synter} is also able to accurately repair 32.4\% of the test cases, which also achieves varying degrees of improvement over CEPROT (4.4\%) and NAIVELLM (28.7\%).

\begin{table}[hptb]
  \footnotesize
  \centering
  \caption{Effectiveness of repairing obsolete test cases caused by SynBCs based on textual match.}
  \label{tab:basicres}
  \begin{center}    
    \begin{tabular}{lcccc}
      \toprule
      \multicolumn{1}{c}{\multirow{2}{*}{\textbf{Approach}}} & \multicolumn{1}{c}{\multirow{2}{*}{\textbf{SPR(\%)}}} & \multicolumn{3}{c}{\textbf{\textit{Textual Match}}}   \\ \cmidrule(r){3-5} 
      \multicolumn{2}{r}{} & \textbf{CodeBLEU} & \textbf{DiffBLEU} & \textbf{Accuracy(\%)} \\
      \midrule
        \textsc{Ceprot}     & 47.8\%    & 73.5     & 26.3     & 4.4\%     \\
        \specialrule{0em}{1pt}{1pt}
        \textsc{Naivellm}   & \textbf{100\%}     & 81.9     & 44.0     & 28.7\%     \\
        \specialrule{0em}{1pt}{1pt}
        \textsc{Synter}  & \textbf{100\%}     & \textbf{83.3}     & \textbf{46.7}     & \textbf{32.4\%}           \\
    \bottomrule
    \end{tabular}
    \end{center}
\end{table}

\subsubsection{Human Evaluation on Effectiveness}
Since textual metrics focus on measuring the similarity of tokens and AST structure for the given codes, they can not well represent the correctness of repair. To bridge this gap, in this part, we replace the obsolete test cases with the generated repaired ones in the repository and manually check whether the replaced test is correct or not, where a correct repair should pass the compilation and keep the intent of the test unchanged.

\begin{table}[t]
  \footnotesize
  \centering
  \caption{Effectiveness of repairing obsolete test cases caused by SynBCs based on intent match.}
  \label{tab:humanres}
  \begin{center}    
    \begin{tabular}{lccc}
      \toprule
      \multicolumn{1}{c}{\multirow{2}{*}{\textbf{Approach}}} & \multicolumn{1}{c}{\multirow{2}{*}{\textbf{CPR(\%)}}} &  \multicolumn{2}{c}{\textbf{\textit{Intent Match}}}  \\ \cmidrule(r){3-4}
      \multicolumn{2}{r}{} &  $\textbf{Repairability}_{\textbf{\textit{gt}}}\textbf{(\%)}$ & $\textbf{Repairability}_{\textbf{\textit{ori}}}\textbf{(\%)}$ \\
      \midrule
        \textsc{Ceprot}     & 33.3\%  & 19.9\% (\textit{27}) & 25.7\% (\textit{35}) \\
        \specialrule{0em}{1pt}{1pt}
        \textsc{Naivellm}   & 88.5\%  & 69.1\% (\textit{94}) & 82.4\% (\textit{112}) \\
        \specialrule{0em}{1pt}{1pt}
        \textsc{Synter}  & \textbf{96.2}\%   & \textbf{75.0\% (\textit{102})} & \textbf{90.4\% (\textit{123})}      \\
    \bottomrule
    \end{tabular}
    \end{center}
\end{table}

The column \textbf{\textit{CPR}} in Tab.~\ref{tab:humanres} represents the success rate of compilation after repairing the original test case in the repository. As a result, \textsc{Synter} fixes the most compilation errors caused by SynBCs. In terms of the two metrics of repairability, \textsc{Synter} both outperforms the baselines. As shown in the column \textbf{\textit{Intent Match}}, we can observe that \textsc{Synter} correctly repair 75.0\% (102/136) and 90.4\% (123/136) test cases in alignment with the intent of the ground truth and the original test respectively.

As for $\text{Repairability}_\text{\textit{ori}}$ specifically, \textsc{Synter} achieves improvements of 248.6\% and 9.8\% when compared to \textsc{Ceprot} and \textsc{Naivellm} respectively. 



\begin{center}
\begin{tcolorbox}[colback=gray!10,
                  colframe=black,
                  width=8.5cm,
                  arc=1mm, auto outer arc,
                  boxrule=0.5pt,
                  left=1pt, right=1pt, top=1pt, bottom=1pt
                 ]
\textbf{Answering RQ1:} \textsc{Synter} outperforms the baselines on all the metrics in the benchmark dataset. It indicates that our approach can effectively help developers to correctly repair obsolete test cases caused by SynBCs.
\end{tcolorbox}
\end{center}

\subsection{RQ2: Effectiveness of TROCtx}
\label{sec:rq2}
As introduced in Section~\ref{sec:intro}, \textsc{Synter} constructs TROCtxs to enhance LLMs, which serves as contextual information to reduce hallucinations during the generation of LLM (LLM below all refers to \textit{GPT-4}). In this research question, we focus on illustrating the effectiveness of TROCtxs. Therefore, we compare our proposed \textsc{Synter} with \textsc{Naivellm} based on the remaining hallucinations.

According to existing work~\cite{hall} and our investigation, we define hallucinations as using undefined methods, variables, or classes in the LLM-generated codes. Based on the results of human evaluation, we manually identify hallucinations and summarize them into two types: common hallucinations and outdated hallucinations.

\textbf{Common hallucinations}. Common hallucinations are caused by the lack of direct contextual information from the change of focal method signature. For the case in Fig.~\ref{fig:case}, a new class (\texttt{MountPOptions}) is set as the third parameter type. Therefore, the LLM falls into common hallucination without the required class contexts of \texttt{MountPOptions}.

\begin{figure}[!ht]
\centering
\setlength{\abovecaptionskip}{5pt}
\setlength{\belowcaptionskip}{-5pt}
\includegraphics[width=0.45\textwidth]{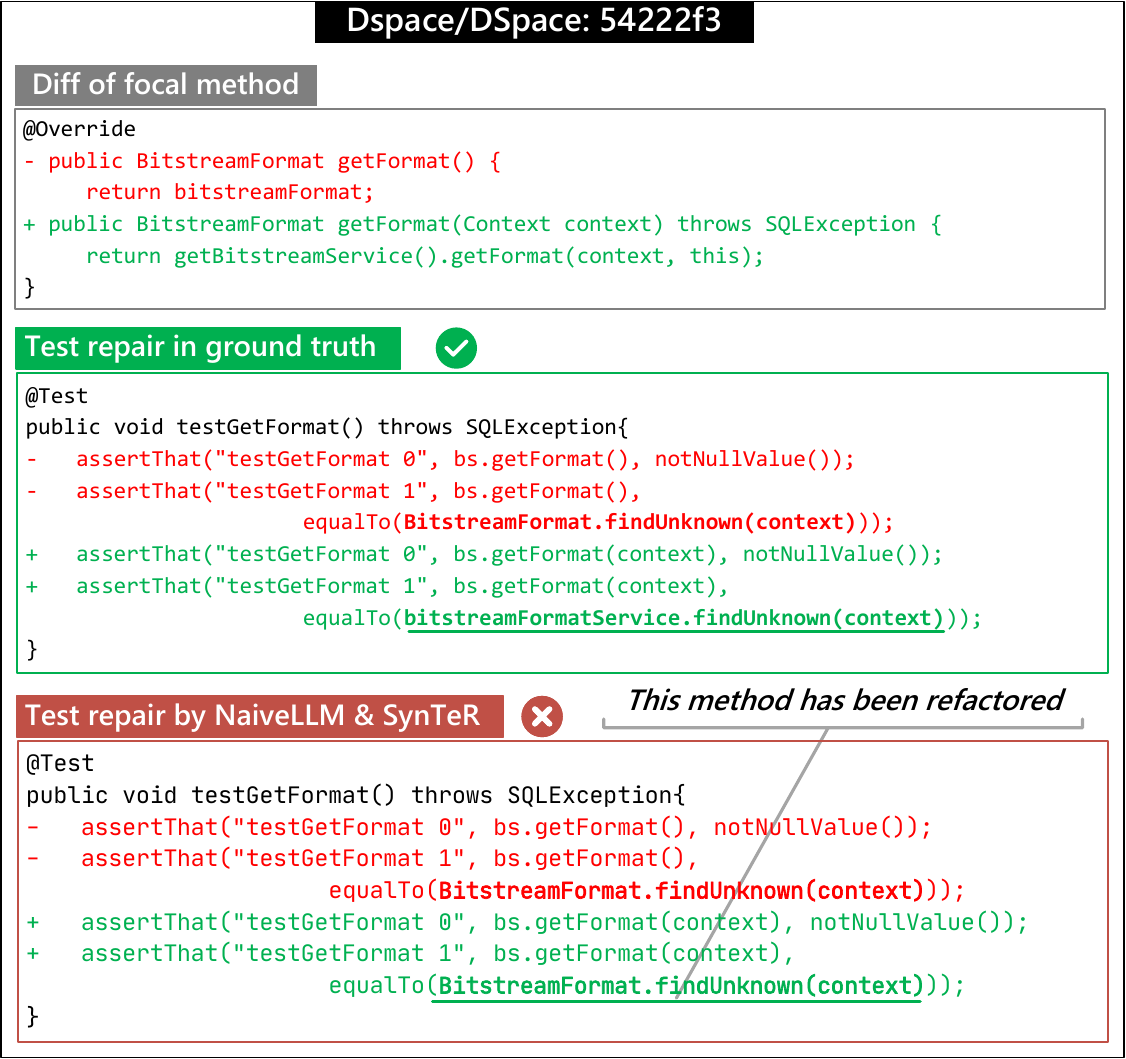}
\caption{An example of outdated hallucination in which LLM incorrectly generates the test with a refactored method.}
\label{fig:failcase}
\end{figure}

\textbf{Outdated hallucinations}. Outdated hallucinations are caused by the lack of implicit contextual information beyond the focal method signature. For the example in Fig.~\ref{fig:failcase}, the method \texttt{findUnknown} used in the test is refactored to be accessed from an instance of class \texttt{BitstreamFormat}. Without this specific knowledge, the test generated by LLM still uses the method in an outdated way, which causes a hallucination.

To assess the effectiveness of TROCtxs constructed by \textsc{Synter}, we collect all the hallucinations for \textsc{Naivellm} and \textsc{Synter}. As shown in Tab.~\ref{tab:hall}, hallucinations occur in both approaches. For the 6 common hallucinations that \textsc{Naivellm} fails, \textsc{Synter} fixes all of them and correctly generates repaired tests with related contexts, indicating that \textsc{Synter} has the general capability to precisely and effectively collect required contexts from the syntactic change of the focal method. Besides, \textsc{Synter} also reduces outdated hallucinations by 25\%, in which two cases are fixed with the constructed TROCtxs (UsageCtxs and EnvCtxs specifically).

\begin{table}[t]
  \small
  \centering
  \caption{The number of different types of Hallucinations.}
  \label{tab:hall}
  \begin{center}    
    \begin{tabular}{lcccc}
      \toprule
      \textbf{Approach} & \textbf{Common Hall.} & \textbf{Outdated Hall.} & \textbf{Total} \\
      \midrule
      \textsc{Naivellm} & 6 & 8 & 14 \\
      \textsc{Synter} & \textbf{0 (-100.0\%)} & \textbf{6 (-25.0\%)} & \textbf{6 (-57.1\%)} \\
      \bottomrule
     \end{tabular}
  \end{center}
  \end{table}
  
\begin{center}
\begin{tcolorbox}[colback=gray!10,
                  colframe=black,
                  width=8.5cm,
                  arc=1mm, auto outer arc,
                  boxrule=0.5pt,
                  left=1pt, right=1pt, top=1pt, bottom=1pt
                 ]
\textbf{Answering RQ2:} The TROCtxs constructed by \textsc{Synter} can effectively reduce the total hallucinations of LLM by 57.1\%, in which all the common ones are fixed.
\end{tcolorbox}
\end{center}
  
\subsection{RQ3: Failure Analysis}
\textsc{Synter} is designed to repair the original test case with its intent unchanged. Considering that the ground truth may contain semantic changes such as adding new assertions and altering input values, we further investigate the 13 cases that fail in terms of $\text{Repairability}_\text{\textit{ori}}$ according to Tab.~\ref{tab:humanres}. Finally, we summarize three reasons for the failure.

\begin{itemize}
    \item \textbf{Uses unimported classes}. For four cases, LLM directly uses classes that are not imported in the file of the test method when generating codes. 
    The key reason is that we focus on collecting contexts by analyzing the syntactic change of the focal method at the method level. Without analyzing the focal class and changes in the methods, the constructed contexts by \textsc{Synter} may be inadequate and result in failure. Fortunately, this type of failure is easy to fix.
    
    \item \textbf{Complex focal changes}.  For eight cases, LLM fails to correctly generate the expected tests with limited contexts as the changes of focal methods in class-level or repo-level upgrades are complex. Specifically, the complex focal change leads to outdated hallucinations and incorrect invocations. This type is challenging to resolve because of the difficulty of collecting implicit contexts and the limited capability of current LLMs. 
    
    \item \textbf{Fails to construct TROCtxs}. Specifically for one case, we find that the language server fails to provide intelligent features for repositories containing configuration errors, which results in the failure of \textsc{Synter} to construct TROCtxs. Overall, it is hard even for developers to manually repair tests in these repositories, in which they can not infer related contexts in IDEs either.
\end{itemize}



\begin{center}
\begin{tcolorbox}[colback=gray!10,
                  colframe=black,
                  width=8.5cm,
                  arc=1mm, auto outer arc,
                  boxrule=0.5pt,
                  left=1pt, right=1pt, top=1pt, bottom=1pt
                 ]
\textbf{Answering RQ3:} \textsc{Synter} fails to repair obsolete tests mainly for using unimported classes, being unaware of complex focal changes, and encountering errors when initializing the language server.
\end{tcolorbox}
\end{center}

\subsection{RQ4: Efficiency}
As shown in RQ1, the performance of \textsc{Ceprot} lags behind other approaches because of the limited model backbone. Although \textsc{Ceprot} can repair tests fast, most of the generated codes are incomplete and contain syntax errors (52.2\%). Therefore, we focus on  comparing the efficiency of \textsc{Naivellm} and \textsc{Synter} in this research question. 

Compared to \textsc{Naivellm}, \textsc{Synter} adds the steps to construct TROCtxs. When querying LLM to repair the test, the prompt of \textsc{Synter} is longer with the constructed TROCtxs. In this research question, we evaluate the time efficiency and the token count of \textsc{Synter} compared to \textsc{Naivellm}. 

\begin{figure}[!ht]
\centering
\setlength{\abovecaptionskip}{5pt}
\setlength{\belowcaptionskip}{-5pt}
\includegraphics[width=0.48\textwidth]{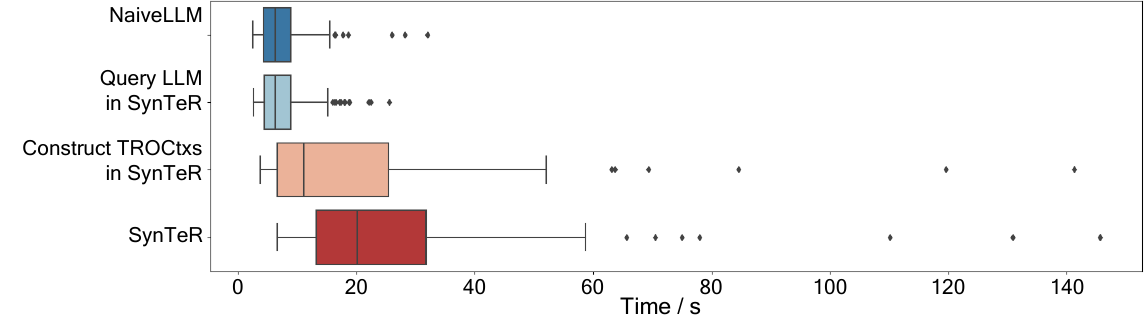}
\caption{The time cost of \textsc{Naivellm} and \textsc{Synter}}
\label{fig:time}
\end{figure}

In particular, we divide the process of \textsc{Synter} into constructing TROCtxs and querying LLM for repair. As shown in Fig.~\ref{fig:time}, the time cost of \textsc{Synter} mainly depends on the process of constructing TROCtxs, while the response time of LLM API of the two approaches is close. We also notice that the language server may fail to synchronize old repositories due to connection timeout for missing dependencies, which results in the outliers with much longer time. In our design, \textsc{Synter} trades time for accuracy. Besides, the average time of \textsc{Synter} falls into 10s to 30s, which is acceptable for developers since test repair is not an interactive task.

\begin{figure}[!ht]
\centering
\setlength{\abovecaptionskip}{5pt}
\setlength{\belowcaptionskip}{-5pt}
\includegraphics[width=0.48\textwidth]{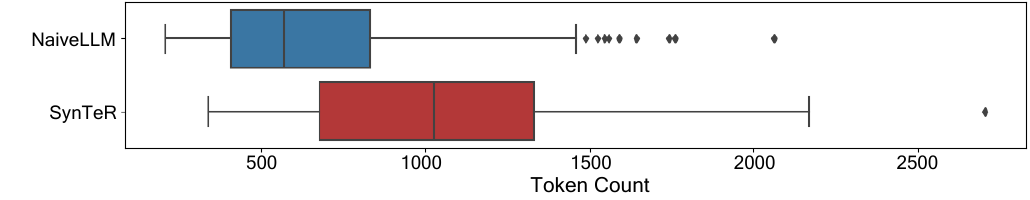}
\caption{The token cost of \textsc{Naivellm} and \textsc{Synter}}
\label{fig:token}
\end{figure}

In terms of the token cost of \textsc{Synter}, we observe that the tokens of the prompt roughly doubled in number after being augmented with TROCtxs from Fig.~\ref{fig:token}. This evidence also indicates that the TROCtxs constructed by \textsc{Synter} are precise considering the large token size of the whole repository.


\begin{center}
\begin{tcolorbox}[colback=gray!10,
                  colframe=black,
                  width=8.5cm,
                  arc=1mm, auto outer arc,
                  boxrule=0.5pt,
                  left=1pt, right=1pt, top=1pt, bottom=1pt
                 ]
\textbf{Answering RQ4:} \textsc{Synter} trades time for accuracy, in which the average time of constructing TROCtxs is 10s to 30s. Besides, the final prompt of \textsc{Synter} contains roughly twice as many tokens as \textsc{Naivellm}.
\end{tcolorbox}
\end{center}

\section{Discussion}

\subsection{Limitations}
The key limitations of \textsc{Synter} exist in its two main modules, the static collector and the neural reranker. First, \textsc{Synter} cannot construct related contexts if the language server fails to initialize due to configuration errors in the repository. Second, the approach's effectiveness is influenced by its reranking strategy. Despite crafting tailored reranking queries for different contexts, we cannot always ensure the precision of reranking. Additionally, our approach focuses on method-level signature changes and does not improve the performance of test repair for class-level or complex implicit changes specifically.

\subsection{Threats to Validity}
\textbf{\textit{External Validity.}} The main threats to external validity come from the evaluation dataset. We reuse the existing dataset, which may not be representative of all possible real-world syntactic changes, so we check and augment the dataset to collect diverse samples from related commits. Besides, it is inevitable to avoid data leakage from popular open-source projects in GitHub. These projects widely use standard design patterns~\cite{bloch2017effective}, which can be learned by LLM during training. Although experimental results show that \textsc{Naivellm} correctly repairs some of the tests with data leakage, it does not affect the effectiveness evaluation of \textsc{Synter}. According to the analysis in RQ2, \textsc{Synter} outperforms \textsc{Naivellm} in cases where \textsc{Naivellm} fails for hallucinations.

\textbf{\textit{Internal Validity.}} In our experiments, a major threat to internal validity is the possible bias in human evaluation. To mitigate it, we invite two senior developers to manually verify the generated tests and annotate explanations. The final result is collected after they reach an agreement after discussion.

\section{Related Work}
\subsection{Production-test Co-evolution}
Production-test co-evolution refers to co-evolving the test codes with the changes in production codes. Most of the previous studies focus on identifying production-test co-evolution pairs~\cite{lubsen2009using},~\cite{hurdugaci2012aiding},~\cite{marsavina2014studying},~\cite{sun2023revisiting}. Recently, more works adopt learning-based techniques to automatically identify obsolete tests~\cite{wang2021understanding},~\cite{sun2023revisiting}. While these works focus on identification only, our approach targets automatically repairing the obsolete test cases directly to relieve the burden of developers.

Two types of obsolete test cases have attracted the attention of developers, one is the GUI-oriented event sequence test case, and the other is the code-oriented method test case. Nowadays, many researchers studied the automated repair of GUI test cases, especially on Android applications~\cite{memon2003regression},~\cite{sun2023revisiting}. However, for code-oriented method test cases, limited studies are focusing on repairing code-oriented method test cases automatically. One biggest difference between them is that the search space for the GUI-oriented test event is more obvious, which can be obtained by traversing the related GUI widget trees, while the code-oriented method test case has a larger search space in the whole repository.

For code-oriented method test case repairing, 
Hu et al.~\cite{ceprot} proposed the first transformer-based approach to update obsolete tests with two stages, identifying and updating, which is the SOTA work in this area. However, it is based on pre-trained models with fewer parameters and lacks contextual information. Compared to this, our work uses larger language models with automatically constructed contexts from the repository.

\subsection{LLM-based Code Generation}
Automated code generation with LLMs brings huge improvements in production efficiency. Repo-level code generation represents the task of generating codes based on a broader context of the repository~\cite{zhang2023repocoder}. It is challenging due to the lack of domain-specific knowledge, which results in hallucination~\cite{liu2023codegen4libs},~\cite{lu2022reacc}. Several studies leverage Retrieval-Augmented Generation (RAG) to improve the performance of LLM in specific code tasks by providing similar codes or results into the query prompt~\cite{zhang2023repocoder},~\cite{nashid2023retrieval}. However, the contexts are retrieved based on simple metrics such as textual similarity without code semantics. 
Recently, some works have focused on improving the capability of LLM by static analysis~\cite{yang2024enhancing},~\cite{agrawal2024monitor}. Specifically, monitor-guided decoding (MGD)~\cite{agrawal2024monitor} is a novel approach to bring IDE-assistance from developers to LLMs to guide the decoding when generating codes, in which IDE-assistance is providing intelligent features by the language server. Compared to this work, we augment LLMs by combining static analysis with RAG to provide more contexts instead of guiding the decoding process. 

\section{Conclusions}
We propose \textsc{Synter}, an LLM-powered approach to automatically repair obsolete test cases caused by syntactic changes of the focal methods. The key idea of \textsc{Synter} is to combine static analysis and neural rerankers to precisely construct test-repair-oriented contexts from the updated repository, which augments the capability of LLM. Experimental results show that \textsc{Synter}'s effectiveness outperforms baseline approaches on both the textual- and intent-matching metrics. Besides, 
with the augmentation of TROCtx constructed by \textsc{Synter}, hallucinations are reduced by 57.1\%.
 The overall results also demonstrate that adopting static analysis techniques to improve the capability of LLM yields excellent performance and could be extended to other code-related tasks.



\section*{Acknowledgements}
Thanks to Dr. Xutong Ma for the initial discussions on this work and to all the anonymous reviewers for their helpful comments and suggestions. This work is supported by the National Natural Science Foundation of China (NSFC) under grant numbers 62132020 and 62102405, and Major Project of ISCAS (ISCAS-ZD-202302).

\bibliographystyle{IEEEtran}
\bibliography{refs} 

\end{document}